\documentclass[prb,twocolumn,english,superscriptaddress]{revtex4-1}

\usepackage[T1]{fontenc}
\usepackage[latin9]{inputenc}
\usepackage{color}
\usepackage{babel}
\usepackage{latexsym}
\usepackage{float}
\usepackage{amsmath}
\usepackage{amsfonts}
\usepackage{graphicx}
\usepackage{times}   %% Times Roman font
\usepackage{esint}
\usepackage{subfigure}
\usepackage{verbatim}
\usepackage[unicode=true,pdfusetitle,
 bookmarks=false,colorlinks=true,citecolor=blue,urlcolor=blue,linkcolor=red]{hyperref}

\makeatletter

\@ifundefined{textcolor}{}
{%
 \definecolor{BLACK}{gray}{0}
 \definecolor{WHITE}{gray}{1}
 \definecolor{RED}{rgb}{1,0,0}
 \definecolor{GREEN}{rgb}{0,1,0}
 \definecolor{BLUE}{rgb}{0,0,1}
 \definecolor{CYAN}{cmyk}{1,0,0,0}
 \definecolor{MAGENTA}{cmyk}{0,1,0,0}
 \definecolor{YELLOW}{cmyk}{0,0,1,0}
}

\@ifundefined{date}{}{\date{}}
\AtBeginDocument{
  
}
\makeatother

\newcommand{\SAVE}[1]{}

\newcommand{\newt}[1]{{{#1}}}

\begin{document}
\renewcommand\abstractname{}

\title{Schwinger Boson Mean Field perspective on emergent spins in diluted Heisenberg antiferromagnets}
\author{Shivam Ghosh}
\affiliation{Laboratory of Atomic And Solid State Physics, Cornell University, Ithaca,  NY 14853, USA}
\author{Hitesh J. Changlani}
\affiliation{Department of Physics, University of Illinois at Urbana-Champaign, Urbana, IL 61801, USA}
\author{Christopher L. Henley}
\affiliation{Laboratory of Atomic And Solid State Physics, Cornell University, Ithaca,  NY 14853, USA}
\date{\today}

\begin{abstract}
Using an adaptation of Schwinger Boson Mean Field Theory (SBMFT) for nonuniform systems, 
we study the nature of low-energy spin excitations 
on the square and Bethe lattice at their percolation threshold. 
The optimal SBMFT parameters are interpreted as on-site potentials 
and pairing amplitudes, which enables an explanation of why emergent local 
moments develop in this system on dilution 
[L. Wang and A. W. Sandvik, Phys. Rev. Lett. 97, 117204 (2006); H.J. Changlani {\it et al., ibid} 
111, 157201 (2013)] and why the corresponding single particle frequencies are driven to \emph{anomalously} low values. 
We discuss how our mean field calculations suggest the strong link between 
the presence of sublattice imbalance and long range antiferromagnetic order, 
and why linear spin wave theory is inadequate for capturing this relation. 
Within the SBMFT framework, we also extract an energy scale for the 
interaction between emergent moments, which show qualitative agreement with many-body 
calculations. 

\end{abstract}

\maketitle

\section{Introduction} The concept of emergence is central to condensed matter systems. This 
means that an effective low-energy description of a system can be made in terms of 
emergent degrees of freedom and interactions between them, 
which might have different properties compared to the original ingredients. 
For example, dilution of quantum magnets in the form of vacancies or substitution with nonmagnetic ions 
creates emergent local moments~\cite{ hoglund,willans, arnab,wangandsandvik,changlani-ghosh} 
which influence the spin texture~\cite{hoglund2}, 
magnetic susceptibility~\cite{hoglund,hoglund3}, specific heat~\cite{hoglund} and 
excitation spectra~\cite{wangandsandvik,changlani-ghosh}. 
Recent work also shows how random on-site magnetic fields in spin chains 
create emergent composite spins which are exponentially localized in space, 
resulting in a non-equilibrated "many-body localized" state~\cite{pal_huse,huse_nandkishore}. 
 
The subject of interest here is that of dilution to the percolation 
threshold i.e. when the nonmagnetic impurities are randomly distributed and their number is 
macroscopically large enough to create a finitely ramified fractal cluster. In this case 
an anomalous energy scale in the low-energy spectrum appears, 
lower than the usual rotor states~\cite{Anderson}. The presence of local regions with an excess of one kind of 
sublattice sites over another give rise to "dangling spins"~\cite{wangandsandvik}, 
whose mutual interactions create low-energy quasidegenerate states~\cite{changlani-ghosh}. 
The energy splittings are exponentially small in the average separation between two such 
emergent spins. 

While many facets of this problem are now understood, a simple explanation 
for the decoupling of such a localized moment~\cite{clh} 
from the rest of the background~\cite{sandvik2002} has remained elusive. 
A mean field explanation at the level of linear spin wave 
theory (LSWT) does not yield meaningful results~\cite{brayali}; the usual 
\newt{N\'eel} state is a not a good starting point, owing to 
the presence of coexisting locally disordered and ordered regions created by dilution. 
Therefore an attractive possibility for \newt{explaining this effect} is the 
Schwinger Boson Mean Field Theory (SBMFT) for quantum antiferromagnets~\cite{arovas,read}, 
which is capable of capturing a wide variety of phases.

The purpose of this paper is thus twofold. 
Our first aim is to demonstrate the utility of SBMFT 
in the context of dilution disorder. 
Going beyond limited functional forms for the mean field parameters, often used for clean systems, 
we instead numerically optimize all the parameters to minimize the energy, 
subject to them satisfying certain constraint equations. 
We find excellent qualitative agreement with respect to 
accurate many-body ground state calculations carried out 
with density matrix renormalization group (DMRG)~\cite{white}.% and quantum Monte Carlo (QMC). 

The second and main aim of the paper is to \emph{interpret} 
the meaning of (1) the parameters corresponding to the lowest energy solution and 
(2) the low lying single particle modes obtained from SBMFT. 
This framework explains the \emph{fundamental} reason for the 
near decoupling of "dangling regions" in a diluted system. 
Our calculations have been carried out for 
\newt{Heisenberg antiferromagnets (HAF)} on the square and 
Bethe lattice site-diluted to their percolation threshold (which 
corresponds to 40.72\% and 50\% dilution respectively). 
While the square lattice case has been extensively studied and is 
relevant experimentally~\cite{vajk_expt}, similar qualitative insights 
have been gained by studying the problem by eliminating loops 
(Bethe lattice)~\cite{changlani-ghosh}. 

The dangling regions weakly interact with each other over the rest of the sites and 
form an effective unfrustrated low-energy system of their own 
that maintains long range order in this system. 
We provide evidence for these assertions by 
studying the single particle spectrum of SBMFT carefully and showing 
the existence of Goldstone modes. These modes are found to significantly 
differ from the corresponding LSWT counterparts; 
the latter is partially improved by the inclusion of quartic terms which 
are treated in a self-consistent Hartree Fock formalism. 
Finally, we connect our SBMFT results with many-body calculations 
and use the numerical spin-spin 
correlators from the theory to obtain a bound on 
the lowest energy scale within a single mode approximation (SMA) formalism.

\section{Schwinger Boson Mean Field Theory (SBMFT) formalism} 
SBMFT~\cite{arovas,read} has been widely successful in capturing a variety of ordered and disordered 
phases of Heisenberg Hamiltonians on regular 
lattices~\cite{sachdev, chung, messio, messiochiral}. In particular, SBMFT has had 
good quantitative agreement with many-body numerical results for systems with 
long range magnetic order~\cite{gazza} and in cases where true \emph{physical} excitations 
of the HAF can be created at the mean field level using SBMFT parameters~\cite{tay}. 
However, only few studies exist where SBMFT or fermionic $\mathbb{SU}(N)$ theories 
have been applied to probe spatially 
nonuniform states~\cite{kaul, hermele, misguich, purebethe}. 
Percolation clusters, with their nonuniform geometry create a natural setting for 
studying spatially inhomogeneous mean field states. 

Here we study the nearest neighbor $S=1/2$ \newt{HAF},
\begin{equation}
\mathcal{H}=\sum_{\langle ij \rangle}J_{ij}\mathbf{S}_{i}\cdot \mathbf{S}_{j}
\label{eq:haf}
\end{equation}
with uniform couplings $J_{ij}=J$ on the square and coordination-3 
Bethe lattice diluted to the percolation threshold. For this Hamiltonian, 
the $\mathbb{SU}(2)$ spin operators are mapped to two \emph{flavors} of Schwinger bosons, 
using the relations, 
\begin{equation}
\mathbf{S}_{i}^{+}=b_{i1}^{\dagger}b_{i2} \;\;\;\; 
\mathbf{S}_{i}^{-}=(\mathbf{S}_{i}^{+})^{\dagger} \;\;\;\;\
2\mathbf{S}_{i}^{z}=(b_{i1}^{\dagger}b_{i1}-b_{i2}^{\dagger}b_{i2}) 
\end{equation}
where $b_{im}^{\dagger}(b_{im})$ is the creation (annihilation) operator for a 
boson of flavor $m$ ($m=1,2$) at site $i$. Once this substitution is performed, 
the resulting Hamiltonian is quartic in the bosonic operators 
and is decoupled by extending the number of flavors from two to $N$ either in the 
$\mathbb{SU}(N)$~\cite{arovas} or $\mbox{Sp}(2N)$~\cite{read} approaches. 

We outline the former approach~\footnote{We also carried out the $\mbox{Sp}(2N)$ formalism and verified 
the results to be consistent with the $\mathbb{SU}(N)$ approach outlined in text}, 
which is valid for bipartite lattices. Within this formalism, the mean field Hamiltonian 
for each boson flavor $m$ is identical and given by,
\begin{equation}
\mathcal{H}_{MF}^{m}= \boldsymbol{\beta}^{\dagger} \left(\begin{array}{cc} \mathbf{\Lambda} & \mathbf{Q}\\ \mathbf{Q}^{\dagger} & \mathbf{\Lambda} \end{array}\right) \boldsymbol{\beta}+\frac{1}{J}\sum_{i<j}|Q_{ij}|^{2}-\left (S+\frac{1}{2}\right)\sum_{i}\lambda_{i}
\label{eq:sbmft ham}
\end{equation}
where $S$ is the spin length, $\boldsymbol{\beta}$ is a vector given by 
$\boldsymbol{\beta}^{T}=(b_{1m},...,b_{N_{s}m},b_{i1}^{\dagger},...,b_{N_{s}m}^{\dagger})$, 
with $N_s$ being the number of sites on the lattice. 

The matrix $\mathbf{\Lambda}$ is diagonal in the site basis 
with entries given by Lagrange multipliers $\lambda_{i}\delta_{ij}/2$ which 
enforce the 'number constraint' on the bosons,
\begin{equation} 
\sum_{m} \langle b_{im}^{\dagger}b_{im}\rangle=NS,
\label{eq:number constraint}
\end{equation} 
which is only satisfied on \emph{average}. This constraint 
maps the Hilbert space of the bosons to that of spins. 
The expectation $\langle ... \rangle$ for evaluating the boson number expectation 
is taken in the Schwinger Boson mean field state.
 
The matrix $\mathbf{Q}$ has entries that are all off-diagonal and are in general 
complex-valued. On loop-less lattices like the diluted Bethe lattice, the \emph{bond variables} $Q_{ij}$ 
can be chosen to be real as there are no non-trivial loop 
fluxes~\cite{messiochiral,misguich} arising from the phases of the bond variables. 
Physically, these parameters denote the strength of the pairing amplitude of bosons; in the spin language 
they denote the strength to form a spin singlet between sites $i$ and $j$. 
The optimal $Q_{ij}$ values in the mean field state satisfy 
\newt{ $Q_{ij} = \langle Q_{ij} \rangle = {(J/N)} \sum_{\langle i,j\rangle,m} \langle b_{im}b_{jm} \rangle$ }, 
where the expectation is again taken in the Schwinger Boson mean field state 
\newt{ and summed over the two flavors $m=1,2$ (for number of flavors $N=2$).} 

\begin{figure*}[htpb]
\centering
\includegraphics[width=\linewidth]{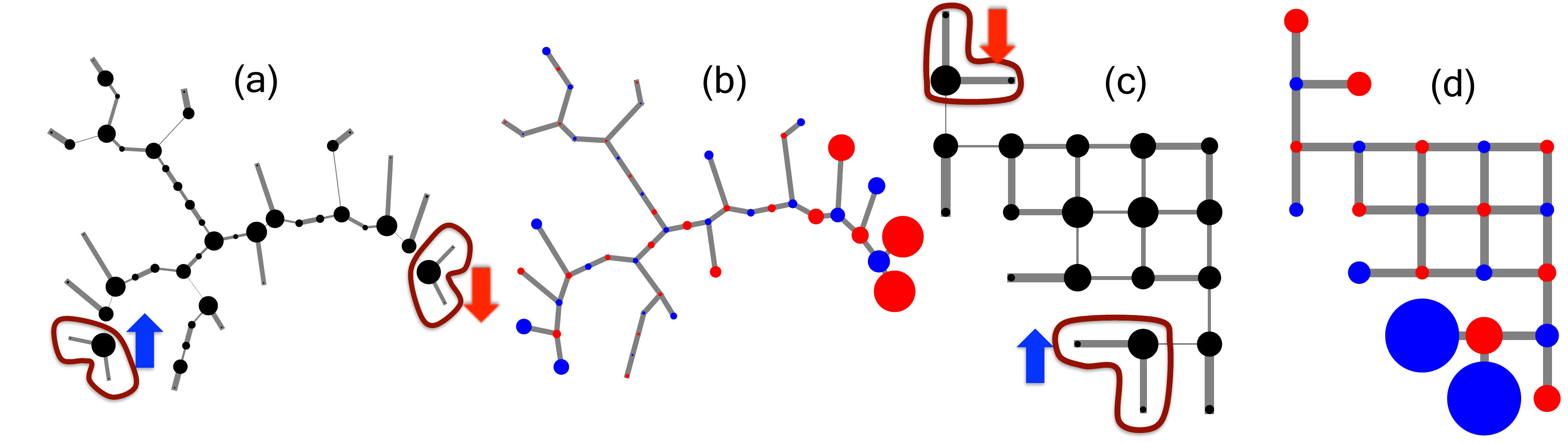} 
\caption{(Color online) Panels (a) and (c) show the optimized SBMFT 
parameters $\lambda_i$, proportional to radius of circles on sites, and 
$|\tilde{Q}_{ij} |-\mbox{min}\{ | \tilde{Q}_{ij} | \}$ for nearest neighbor bonds, 
proportional to the thickness of the bonds, on a diluted Bethe and square lattice respectively. 
Panels (b) and (d) show the lowest single particle eigenmode $\psi^{+}_{i0}$, whose amplitude is 
proportional to the radius of the circles and sign is denoted by the color. 
For more details about the interpretation of the parameters and eigenmodes refer to the text.}
\label{fig:fig1}
\end{figure*}

In general, the theory allows for extended-neighbor mean field parameters (i.e. any pair $i,j$), but 
in this paper we retain nonzero $\{Q_{ij}\}$ corresponding only to nearest neighbors. 
In practice, this restriction is generally found to give solutions that 
qualitatively match results from many-body calculations.

The set of mean field parameters $\{ \lambda_{i},Q_{ij} \}$, collectively called 
an \emph{ansatz}, completely specifies the solutions of SBMFT. They 
are determined variationally by minimizing the mean field energy 
$\langle \mathcal{H}_{MF}^{m} \rangle=e_{MF}^{m}$, subject to constraints. 
$e_{MF}^{m}$ is obtained by first solving the eigenvalue problem~\cite{colpa}, 
\begin{equation}
\left(\begin{array}{cc} \mathbf{\Lambda} & \mathbf{Q}\\ -\mathbf{Q}^{\dagger} & -\mathbf{\Lambda} \end{array}\right) \left( \begin{array}{cc} u_{n} \\ v_{n} \end{array}\right)=\omega_{n}\left( \begin{array}{cc} u_{n} \\ v_{n} \end{array}\right)
\label{eq:eigenvalue eq}
\end{equation}
which gives the single particle Bogoliubov modes labeled $n$ 
with frequencies $\{ \omega_{n}\}$. This diagonalization gives 
$2N_{s}$ frequencies which occur as 
$N_{s}$ $\pm |\omega_{n}|$ pairs. It is only the 
positive set of frequencies that are relevant for 
the quasiparticle spectrum. The wave functions corresponding to these frequencies 
are denoted as $\{ u_{in},v_{in} \}$. 
A linear combination of the modes $u_{in},v_{in}$ is 
taken to define a length $N_{s}$ mode on the lattice, 
\begin{equation} 
\psi_{in}^{\pm} = u_{in}\pm v_{in}.
\label{eq:psiin}
\end{equation}
where $\psi_{in}^{\pm}$ are eigenvectors of an $N_{s}\times N_{s}$ matrix with eigenfrequencies $\{\omega_{n}^{2}\}$; 
\newt{more} details of this matrix have been discussed elsewhere~\cite{mucciolo}. 
For the lowest frequency SBMFT modes, we found $u_{in}$ ($v_{in}$) 
to be zero (nonzero) on one sublattice and nonzero (zero) 
on the other. \newt{For such modes, we fix the choice of gauge in the definition of $\{Q_{ij}\}$ 
by defining all values to be positive or negative, so that the wave function has a staggered sign pattern. 
The resultant mode, which we refer to as $\psi_{in}$, is then used to compute all further operator expectations. 
This choice of gauge is completely equivalent to a choice of a uniform sign pattern 
and does not change the expectations of any physical observables.} 

The zero-point quantum energies $\{ \hbar\omega_{n}/2\}$ are summed 
to get $e_{MF}^{m}$ (the mean field energy per flavor), 
\begin{equation}
e_{MF} = \sum_{n=1}^{n=N_{s}}\frac{1}{2}\omega_{n}+ e_{classical}
\label{eq:energy per flavor}
\end{equation}
where equivalence between flavors allows for the dropping of the flavor 
index $m$ and $e_{classical}={(1/J)}\sum_{\langle ij \rangle}Q_{ij}^{2}-\left(S+\frac{1}{2}\right)\sum_{i=1}^{i=N_{s}}\lambda_{i}$.
Connection with the \emph{physical} Heisenberg spins is made for $N=2$ and the 
energy for this special case is given by $E_{Heis}=2e_{MF}+\sum_{ij}J_{ij}S^{2}$.

The optimal \emph{ansatz} \newt{ $\{ \tilde{\lambda}_{i}, \tilde{Q}_{ij} \}$ } satisfies the 
'number constraint' and the 'bond constraint': $\tilde{Q}_{ij}=\langle \tilde{Q}_{ij} \rangle$, which 
are implemented by introducing cost functions,
\begin{equation} 
C_{\lambda}\equiv\sum_{i}(\langle b_{i}^{\dagger} b_{i}\rangle-S)^{2} \;\;\; C_{Q} \equiv \sum_{\langle ij \rangle}(Q_{ij}- \langle Q_{ij}\rangle)^{2}
\label{eq:constraints}
\end{equation}
which are made as small as possible. The constrained optimization of the 
mean field variables $\{\lambda_{i}, Q_{ij} \}$ is thus 
transformed to a minimization problem through the cost functions $C_{Q},C_{\lambda}$, which we perform 
efficiently using the Levenberg-Marquardt algorithm~\cite{misguich,levmar}. 
Typically, we found that 
these costs for the solutions reported are in the range of $10^{-19}-10^{-16}$ 
(note the square root is above machine precision). 

\newt{The geometry of the diluted Bethe lattice allows for a simplification 
in the initial choice of the bond amplitudes $\{Q_{ij}\}$, thereby improving the speed and 
scalability (to larger system sizes) of the optimization algorithm. Specifically, the lack of loops on the diluted 
Bethe lattice implies the absence of fluxes and this allows us to choose all initial $\{Q_{ij}\}$ to be real. 
The computational complexity of the algorithm on the Bethe lattice scales as $\sim \tau_{Q}\tau_{\lambda}N_{s}^{3}$, 
where $\tau_{\lambda}(\tau_{Q})$ is the number of optimization steps to minimize $C_{Q}(C_{\lambda})$ and $N_{s}^{3}$ 
is the complexity of diagonalizing the mean field Hamiltonian \eqref{eq:sbmft ham} once. 
However, for the diluted square lattice the bond amplitudes $Q_{ij}$, in general, 
can be complex-valued. This leads to $\text{U}(1)$ fluxes $\Phi$ through even-length 
loops on the lattice, defined in the following gauge invariant manner,
\begin{equation}
\Phi= Q_{ij}(-Q_{jk}^{*})Q_{k\ell}...(-Q_{pi}^{*})
\label{eq:u1flux}
\end{equation}
where $Q_{jk}^{*}$ refers to the complex-conjugate of $Q_{jk}$. The smallest non-trivial even length 
loop on the square lattice is a square plaquette. 

The optimization algorithm is started by allowing 
all initial $Q_{ij}$ to be complex and both the real and imaginary parts of the bond amplitudes are updated 
at every step of the self-consistent cycle to minimize $C_{Q}$ \eqref{eq:constraints}. Since the number of effective 
constraints entering $C_{Q}$ double (real and imaginary part for each bond), the computational cost 
of optimization is roughly twice that of the diluted Bethe lattices~\footnote{In practice, 
the \text{U}(1) gauge invariance of the flux Eq.~\eqref{eq:u1flux} 
can be exploited to set a certain number of bond amplitudes to be purely real without 
changing the flux structure of the mean field state~\cite{misguich}. Hence, in practice, 
the computational cost of the optimization increases only 
by a factor between 1 and 2 compared to the Bethe lattice case}.

The optimal mean field {\it ansatz} $\{ \tilde{\lambda}_{i}, \tilde{Q}_{ij} \}$ on diluted square lattices 
always expels fluxes such that the optimal state has zero flux through all even length loops on the lattice. 
This was verified by starting the optimizer with several initial distributions of bond amplitudes which threaded 
nonzero fluxes through loops on the lattice. As the optimization proceeded, the flux pattern of the state was 
tracked and in all cases we found the ground state to be a zero flux state. 

Finally, we remark that we studied specific instances of both kinds of 
clusters for generating insights and confirming our assertions. However, 
all analyses involving disorder-averaging were studied only for the Bethe lattice case. 
}

\section{SBMFT parameters and single particle modes} 
The interplay between the various contributions to 
$e_{MF}$ in eq.~\eqref{eq:energy per flavor} can be understood heuristically. 
For a uniform one dimensional chain of length $L$~\footnote{Similar functional 
dependencies of $\omega_{n}$ on $\lambda$ and $Q$ are found for small nonuniform geometries} 
the frequency $\{ \omega_{n} \}$ for $\{ \lambda_{i}=\lambda, Q_{ij}=Q \}$, 
momentum $k_{n}$ and coordination $z$ are given by $\omega_{n}=\sqrt{\lambda^{2}-(zQ\cos(k_{n}))^{2}}$~\cite{arovas}. 
For $zQ/\lambda <1$ we have $\omega_{n}\sim \lambda-cQ^{2}/\lambda$ 
($c$ absorbs the momentum dependence) implying that $\omega_{n}$ 
is minimized when $Q^{2}$ is maximized and $\lambda$ is minimized. 
On the other hand, the second and third terms i.e. "classical terms" 
in~\eqref{eq:sbmft ham} favor the opposite i.e. low $Q^{2}$ and high $\lambda$.
This competition between contributions to the energy can be complicated, 
especially in the case of a disordered system, and thus demands a numerical optimization.

Our results for the optimal \emph{ansatz} for representative 
Bethe and square lattices at percolation are shown 
in Fig.\ref{fig:fig1}~(a) and (c) respectively. 
In both figures, the thickness of the bonds is proportional 
to \newt{ $| \tilde{Q}_{ij} |-\mbox{min}\{| \tilde{Q}_{ij}  | \}$ } and the radius of the 
disc on every site is proportional to $\lambda_{i}^{*}$ at that site. 

The distribution of optimal $\{ |\tilde{Q}_{ij} | \}$ in Fig.\ref{fig:fig1}(a) and 
Fig.\ref{fig:fig1}(c) is a prescription for identifying pairs of spins 
with the strongest spin-spin correlations. Since the nearest neighbor 
spin correlations are proportional to the pairing 
amplitudes $\langle \mathbf{S}_{i}\cdot \mathbf{S}_{j} \rangle =3 \tilde{Q}_{ij}^{2} /2$, 
the mean field ground state exhibits strong dimerization (pairing of nearest neighbor spins into singlets) 
as was predicted in DMRG calculations~\cite{changlani-ghosh}. Dimerizing nearest neighbor 
spins for the strongest \newt{$|\tilde{Q}_{ij}|$} bonds pairs up all but two spins on each of the clusters 
in Fig.\ref{fig:fig1}(a) and Fig.\ref{fig:fig1}(c). 

The distribution of $\{ \tilde{\lambda}_{i} \}$ is proportional to the local coordination of the site; 
singly coordinated sites have small $ \tilde{\lambda}_{i}\sim 0.5$ 
and triply coordinated sites have large $ \tilde{\lambda}_{i} \sim 2.5$. 
The $\{ \tilde{\lambda}_{i} \}$ field acts like an on-site disordered potential 
for the bosons. \newt{This can be observed in the 
low frequency wave functions $\psi_{in}$ given by Eq.~\eqref{eq:psiin},}
which avoid sites with high values of the potential, as shown in Fig.\ref{fig:fig1}(b),(d). 
The radius of the discs in \ref{fig:fig1}(b) and (d) are proportional to wave function amplitudes 
and the sign is encoded in the red (positive) and blue (negative) 
colors. Bosons have the highest amplitude of being on sites with the lowest potentials. 

%Each of the two  on the two clusters are found in regions of three 
%sites encircled in Fig.\ref{fig:fig1}(a),(c). 
The two lowest frequency modes are each localized on "fork" regions of three sites, 
(encircled in Fig.~\ref{fig:fig1}(a) and (c)) and decay exponentially in 
to the cluster (Fig.~\ref{fig:fig2}(a)). An exponential fit to the mode on 
the Bethe lattice gives a decay constant $\xi_{\ell oc}$ of about $3$ lattice spacings. 
These localized wave functions are the SBMFT characterization of \emph{emergent} dangling spins on the cluster 
and are completely analogous to similar mode profiles obtained in many-body calculations~\cite{changlani-ghosh}. 

The association of dangling spins with exponentially localized modes is a generic 
feature of both Bethe and square lattice percolation clusters. We check this by using 
a geometrical algorithm~\cite{changlani-ghosh} by isolating $50$-site \newt{Bethe lattice} 
clusters with two dangling forks and fitting exponentially decaying functions of the distance away from the fork tip 
to the two lowest energy modes. The fits give a disorder averaged $\langle \xi_{\ell oc} \rangle_{dis.} \sim 3$ 
lattice spacings in good agreement with the decay length of effective interactions~\cite{changlani-ghosh}. 
This agreement is also indicative of the fact that the 
strength of effective interactions is proportional to the spatial overlap of the two modes. 

\section{Nonuniform Goldstone modes in SBMFT} 
\subsection{Number of modes and effects of intermode interactions} 
The correspondence between dangling spins and localized 
bosonic modes in Fig.\ref{fig:fig1} strongly suggests that each dangling spin leaves its 
own characteristic signature in the SBMFT single particle spectrum $\--$ a low frequency and an 
associated localized mode. This implies that the count of low energy frequencies in the single particle 
spectrum must match the number of dangling spins on the cluster. 

We carry out a systematic check of this assertion by taking an ensemble of clusters and 
deploying techniques developed previously~\cite{changlani-ghosh} to filter out the low energy 
single particle spectrum $\{\omega_{\ell ow} \}$ for each cluster in the ensemble. 
The count of frequencies in $\{ \omega_{\ell ow}\}$ is then tallied against the number of dangling spins 
on a cluster, determined using a geometrical algorithm~\cite{changlani-ghosh}. The situation is complicated 
by the fact that these emergent spins are not totally decoupled; their interactions push certain 
single particle frequencies to higher energies and some to much lower energies. Thus the counts are 
found to agree in $\sim 92\%$ of cases. Part of this discrepancy also arises from spatially extended dangling regions, 
which effectively leads to enhanced interactions with other localized dangling spins. 

\begin{figure}[htpb]
\centering
\subfigure[]{\includegraphics[width=0.95\linewidth]{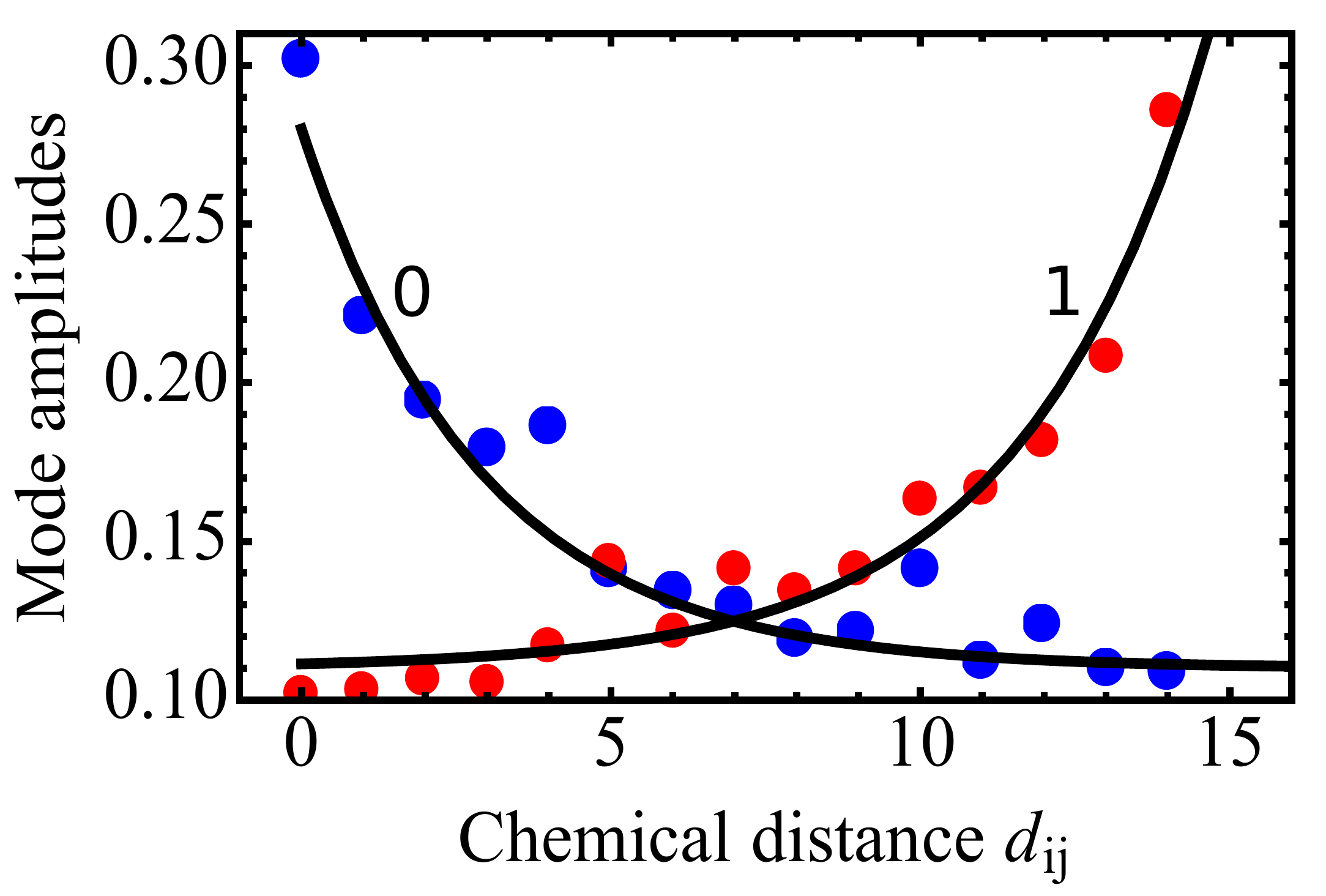}}
\subfigure[]{\includegraphics[width=\linewidth]{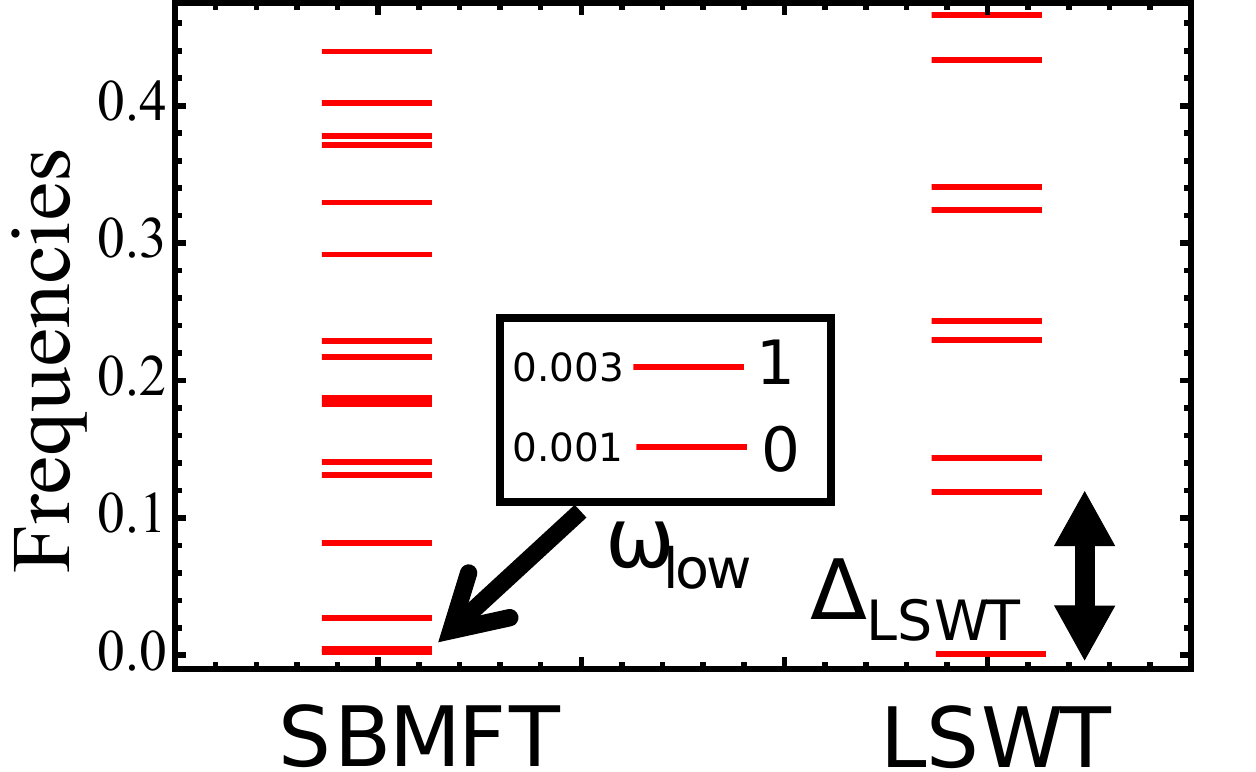}} 
\caption{(Color online)(a) Mode amplitudes $\psi_{in}$ for the two lowest 
frequency single particle modes $n=0,1$ for the \newt{Bethe lattice} percolation cluster 
shown in Fig.~\ref{fig:fig1}(a),(b). Each mode (blue or red points) 
is plotted as a function of separation from the dangling spin at the fork tips. \newt{The fitted 
exponential decay is shown by solid black lines.} Panel (b) shows single particle frequencies 
for the same cluster obtained within SBMFT and LSWT. The two \emph{anomalously} low Goldstone frequencies, 
are indicated by an arrow and labeled as $\omega_{\ell ow}$; \newt{ their numerical 
values and mode numbers ($n$) are indicated in the inset.} The lowest nonuniform spin wave frequency 
is indicated by $\Delta_{LSWT}$.} 
\label{fig:fig2} 
\end{figure}

Among this set of low frequencies $\{\omega_{\ell ow}\}$, we found that two of them were driven to 
\emph{anomalously} low values. These two \emph{anomalously} low frequencies are shown for the Bethe percolation cluster 
of Fig.\ref{fig:fig1} in Fig.\ref{fig:fig2}(b). The $\mbox{SU}(2)$ invariance of the SBMFT formalism, 
along with the fact that these calculations are done on a finite cluster 
prevents these two frequencies from becoming exactly zero. However 
in the thermodynamic limit or in the presence of a small magnetic field, these \emph{anomalously} 
low frequencies will be the first to become zero causing bosons to condense in to this mode. 
This signals long range order within SBMFT and allows identification of the two \emph{anomalously} 
low frequencies as the finite size manifestation of Goldstone modes on the cluster. 

\begin{figure}[htpb]
\centering
\includegraphics[width=\linewidth]{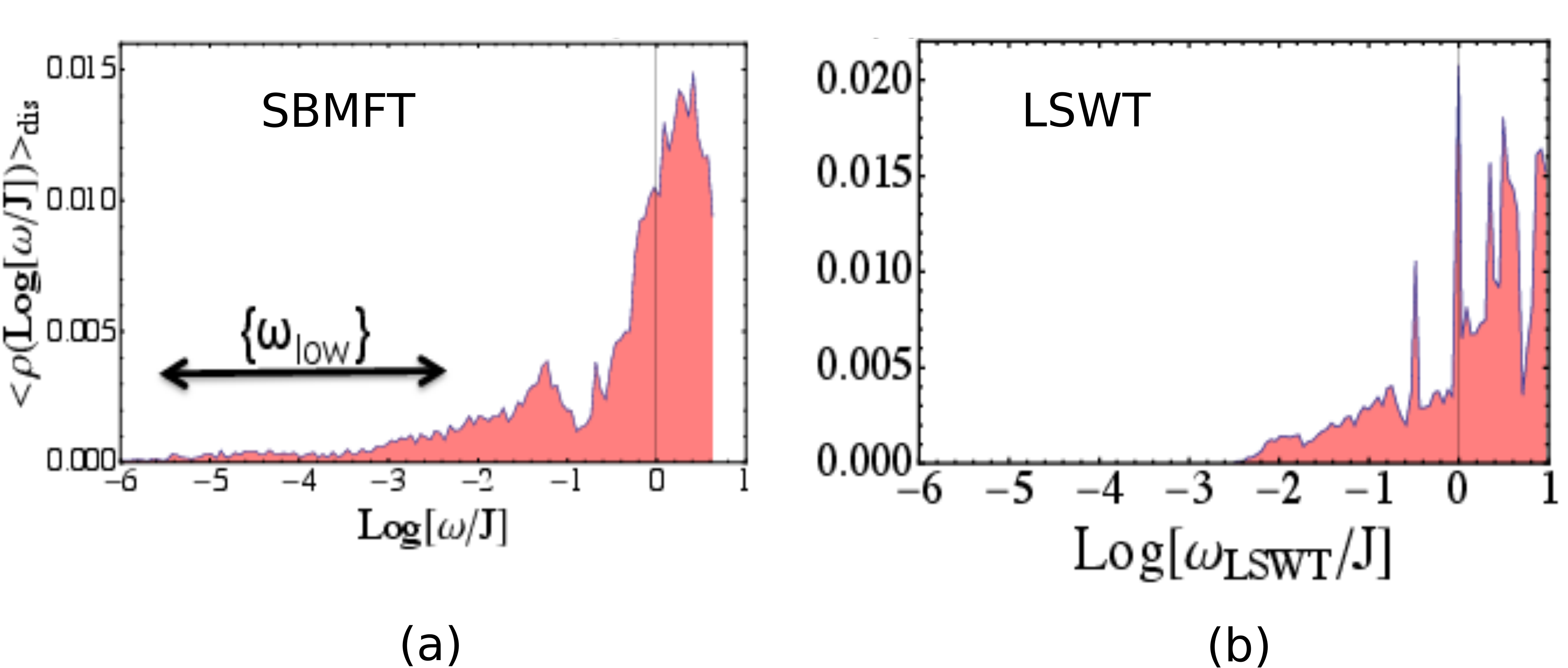}
\caption{(Color online) \newt{ Disorder-averaged single particle 
density of states within (a) SBMFT and (b) LSWT frameworks. An 
ensemble of $400$ Bethe lattice percolation clusters, each 
consisting of 50 sites was considered. The histograms were generated by grouping 
frequencies in bins of size $0.01 J$. The SBMFT calculations show a set of low lying 
frequencies missing in the LSWT spectrum, which have been 
labeled as $\omega_{\ell ow}$.}
}
\label{fig:dos} 
\end{figure}

\subsection{Failure of linear spin wave theory} 
The Goldstone modes, as seen in Fig.~\ref{fig:fig1}(b),(d), have nonuniform amplitudes; 
significantly different from the zero energy uniform Goldstone modes seen in 
LSWT~\cite{mucciolo}. These results suggest that SBMFT, 
in a single unified framework, characterizes emergent 
dangling degrees of freedom by associating localized modes with each of them, 
along with correctly predicting a background of long range \newt{N\'eel} order~\cite{sandvik2002}. 
The maximal amplitude of Goldstone modes on the dangling sites provides direct evidence 
for the crucial role of dangling spins in stabilizing long range order on the cluster. 
\newt{ Based on these insights, we develop a better understanding for 
why LSWT fails to qualitatively capture the nonuniformities 
associated with the Goldstone modes.} 

The crucial difference between the LSWT and SBMFT approaches is that the former 
breaks spin rotational symmetry leading to the inability to capture the 
\emph{anomalous} lowering of frequencies associated with emergent $\mbox{SU}(2)$ 
invariant dangling spin excitations. A comparison between LSWT and SBMFT single 
particle frequencies in Fig.~\ref{fig:fig2}(b) for the Bethe percolation cluster 
in Fig.~\ref{fig:fig1}(a),(b) shows that the lowest LSWT frequency 
is much higher than the corresponding frequency within SBMFT 
(LSWT has two exactly zero frequency uniform amplitude modes by construction). 
We generalize this observation by taking an ensemble of $50$-site Bethe percolation 
clusters and plotting the disorder-averaged density of states (DOS) $\langle \rho(\text{Log}[\omega/J])\rangle_{\text{dis.}}$, $\langle \rho(\text{Log}[\omega_{LSWT}/J]) \rangle_{\text{dis.}}$ within SBMFT 
and LSWT, respectively. 
The DOS on a \newt{logarithmic} scale, calculated within LSWT and SBMFT, 
is shown in Figs.\ref{fig:dos}(a),(b) respectively. The presence of 
an additional low energy scale is seen in the SBMFT DOS and 
is indicated by a probability distribution with a long tail labeled $\omega_{\ell ow}$. 

This discussion motivates us to look closely at the connection between the two approaches.
We note that the LSWT Hamiltonian maps \emph{exactly} to a SBMFT Hamiltonian with \emph{fixed} $\lambda_i$ and $Q_{ij}$. 
In LSWT, $\lambda_i$ equals the coordination of the site, but in 
SBMFT this variable is a degree of freedom that is optimized. 
Since the optimal SBMFT solution is found to deviate 
significantly from the corresponding 
LSWT prediction, we conclude that it is this 
variational freedom that allows the lowest energy modes to have nonuniform amplitudes, 
with largest weights in dangling regions.

\begin{figure}[htpb]
\centering
\includegraphics[width=\linewidth]{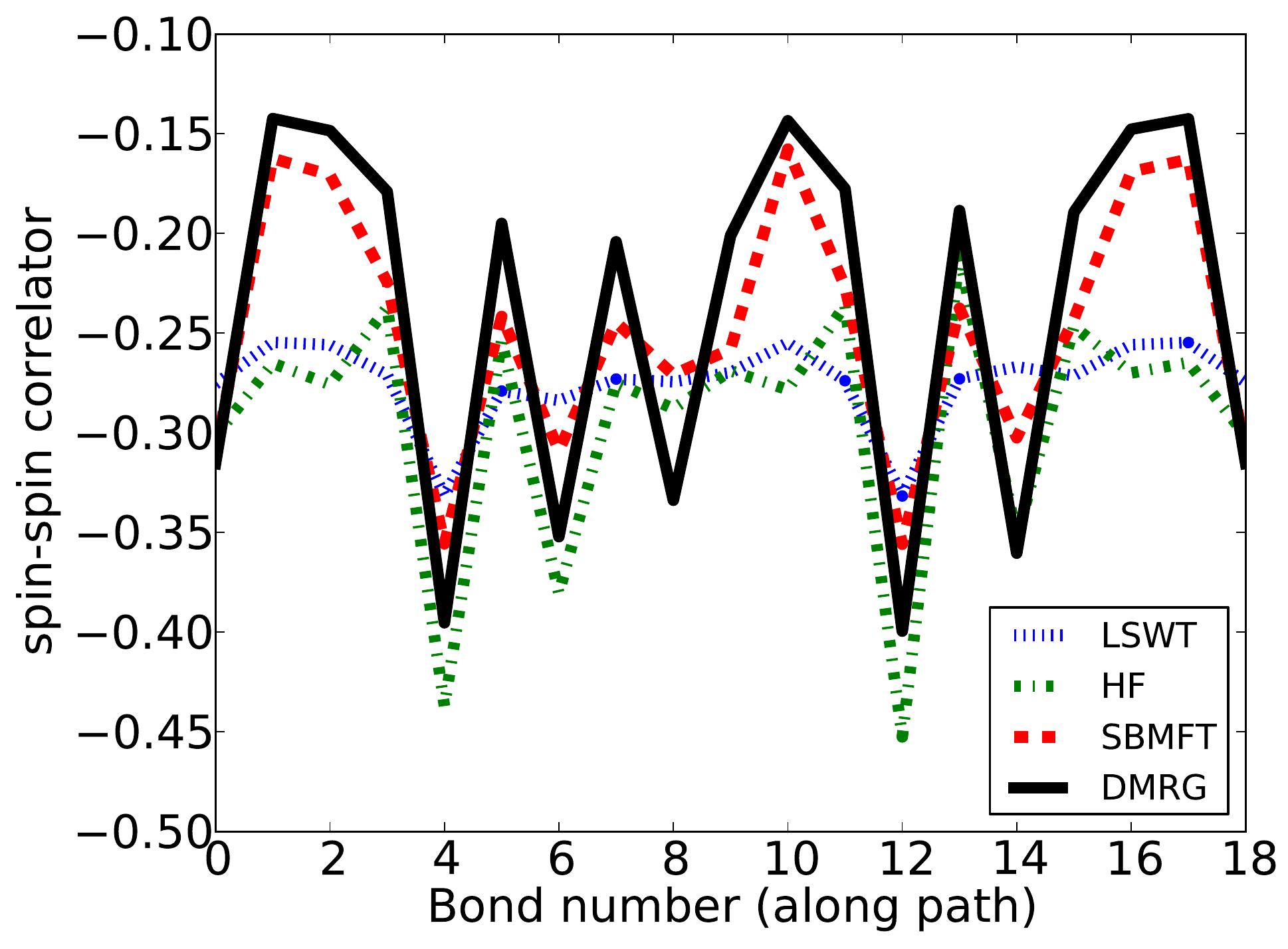} 
\caption{(Color online) Nearest neighbor in plane 
spin-spin correlations, defined to be  $\langle S_i^{x}S_j^{x} + S_i^{y}S_j^{y} \rangle$, 
for bonds taken along the path shown in Fig.~\ref{fig:fig3}(a),
from various methods: linear spin wave theory (LSWT), Hartree Fock (HF), 
Schwinger Boson mean field theory (SBMFT) and the density matrix 
renormalization group (DMRG). The HF results are used from the iteration 
before the convergence of the self-consistency cycle failed. 
LSWT does not capture the variations in the values 
of the correlation functions, while going to order 1/S (S is the spin length) using HF shows 
tendency to capture the behavior of SBMFT and DMRG.}
\label{fig:compare_corrs} 
\end{figure}

Evidence for the \emph{tendency} to form localized modes can be seen by 
including higher order terms in the spin wave expansion; for the 
purpose of this paper we have retained the order $1/S$ terms. 
A Hartree Fock (HF) decoupling of the quartic terms 
is applied and the resultant mean field equations are solved self consistently. 
Every iteration of our calculation cycle lowered the energy and showed gradual 
localization of the lowest modes, but we were unable to converge our solutions. 
This is not a problem with our implementation; instead it is 
evidence of growing spin fluctuations which eventually 
violate the assumptions of the Holstein-Primakoff expansion. 
The general inadequacy of LSWT and partial improvement with HF 
compared to SBMFT is confirmed with our results 
for certain spin-spin correlators shown in Fig.~\ref{fig:compare_corrs}, 
where the three methods are compared to the corresponding near-exact 
DMRG values. (For a detailed exposition of the 
calculation of transverse correlations in spin wave 
theory, we refer the reader to Ref.~\onlinecite{uzi}.) 

%In the next section, we make further assessments 
%of the SBMFT correlators with respect to many-body calculations. 

\section{SBMFT correlators and effective interactions between emergent spins}

\subsection{Comparison of SBMFT and exact calculations}
Through our computations for percolation clusters, we 
have shown that SBMFT can capture the correct qualitative physics of 
disordered systems. %and provides the much needed qualitative insight lacking previously.
The next step is to assess the accuracy of the method with respect to many-body calculations, 
as a means of establishing the legitimacy of our conclusions. A useful product of such comparisons 
is a better evaluation of SBMFT as a computational tool in situations where 
performing accurate many-body calculations may be difficult.

\begin{figure*}[htpb]
\centering
\includegraphics[width=\linewidth]{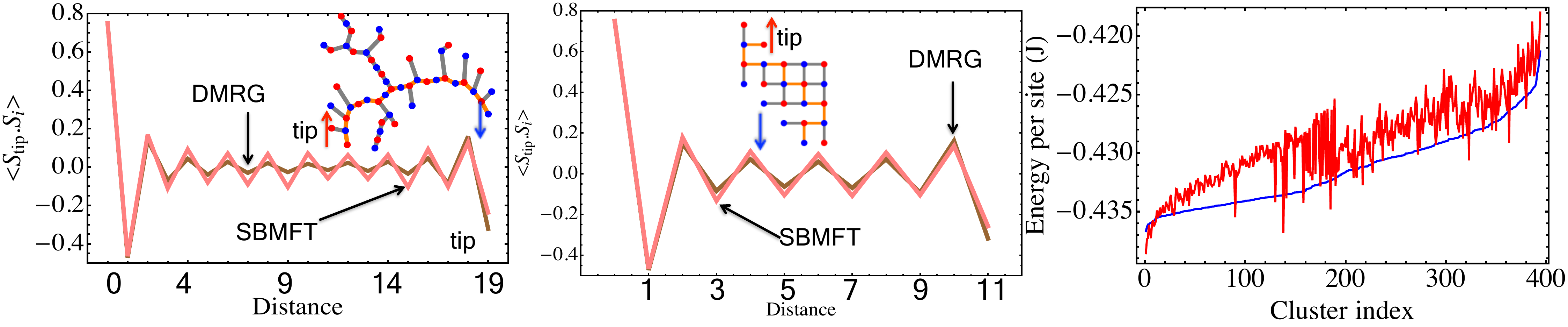} 
\caption{(Color online)(a) Spin-spin correlations comparison between SBMFT and exact DMRG 
results for Bethe cluster of Fig.\ref{fig:fig1}(a),(b), also shown in the inset. 
The correlations are between the tip spin labeled 'tip' and all spins along the path shown in 
orange on the cluster (b) Similar comparison of correlations as 
in (a) for the square lattice cluster in Fig.\ref{fig:fig1}(c),(d). 
(c) Comparison between ground state energies (sorted by value) 
from SBMFT (red):$e_{gs}^{SBMFT}$\eqref{eq:sbmftgs} and DMRG (blue) 
for an ensemble of $400$ size $50$ clusters. The SBMFT energy 
is obtained by summing over nearest neighbor spin correlations. 
}
\label{fig:fig3}  
\end{figure*}

In Figs.~\ref{fig:fig3}(a) and ~\ref{fig:fig3}(b), for a diluted Bethe and square lattice respectively, 
we compare the DMRG 
and SBMFT spin-spin correlators for a reference 
site at the tip of the cluster and the 
other sites along a particular path. While we observe great qualitative agreement, 
quantitatively our SBMFT calculation predicts slightly 
exaggerated oscillations at long distances. 

A metric for a quantitative comparison of SBMFT versus DMRG is the ground 
state energy per site obtained within SBMFT by summing over nearest neighbor spin-spin correlations, defined as:
\begin{equation}
e_{gs}^{SBMFT} = \frac{J}{N_s} \sum_{\langle i,j\rangle} \langle \mathbf{S}_{i}\cdot \mathbf{S}_{j} \rangle
\label{eq:sbmftgs}
\end{equation}
where the expectation $\langle ...\rangle$ is taken in the SBMFT ground state. The accuracy of 
the nearest-neighbor spin-spin correlations found within SBMFT as seen 
in Fig.\ref{fig:fig3}(a) and \ref{fig:fig3}(b), 
ensures that the maximum error in $e_{gs}^{SBMFT}$ compared to the true ground state energy from DMRG, 
for an ensemble of $400$ clusters of $50$-site clusters, is about $1\%$. A comparison of these 
estimates for our ensemble of clusters is shown in Fig.\ref{fig:fig3}(c). 
We also note that $e_{gs}^{SBMFT}$ serves as a better approximation of the ground state energy 
of this model compared to the mean field energy $e_{MF}$. \newt{However, it is important to 
clarify that $e_{gs}^{SBMFT}$, just like $e_{MF}$, is non-variational. The mean field state, 
that is used to compute both these energies, satisfies the boson number constraint \eqref{eq:number constraint} 
only on average}. 

\subsection{Energy scale of effective interactions}
SBMFT is primarily a tool to study singlet ground states, with no direct way
to extract excited state information. Here we use the ground-state 
SBMFT correlators to estimate the singlet-triplet gap and hence the couplings 
between two weakly interacting emergent spins. This estimation is done within 
the single-mode approximation (SMA) formalism~\cite{SMA}, which we briefly explain below. 
An alternate, but related, analysis maximizing the overlap 
of the SMA wave function with the true one (from DMRG) was 
also used previously by us~\cite{purebethe}. 

In the SMA, the triplet excited state $| \boldsymbol{\Psi}_{SMA}\rangle$ is created 
by taking a weighted superposition of single spin excitations of 
the ground state SBMFT wave function, a singlet state ($S^{tot}=0,S_{z}^{tot}=0$), 
\begin{equation}
| \boldsymbol{\Psi}_{SMA}  \rangle= \sum_{i=1}^{N_{s}} w_{i} S_{i}^{+}|  \boldsymbol{\Psi}_{MF}  \rangle
\label{eq:smastate}
\end{equation}
where $\{ w_{i}\}$ are variational weights determined by minimizing the SMA gap 
\begin{equation}
\triangle_{SMA}= J \frac{\langle \boldsymbol{\Psi}_{SMA}| \sum_{\langle ij \rangle} \mathbf{S}_{i} \cdot \mathbf{S}_{j} | \boldsymbol{\Psi}_{SMA} \rangle} {\langle \boldsymbol{\Psi}_{SMA}| \boldsymbol{\Psi}_{SMA} \rangle} - E_0 
\end{equation}
with $E_0$ being the ground state energy. 
 
For the Heisenberg model with uniform bond strengths, assuming a singlet ground state, an 
expression of the gap was derived previously~\cite{purebethe},
\begin{equation}
	\Delta_{SMA} = \frac{- J \sum_{\langle k,l \rangle} (w_{k}-w_{l})^{2}G_{kl}} {2 \sum_{i,j} w_{i}\;w_{j}\;G_{ij}}
\end{equation}
where $\langle k,l \rangle$ are connected links, here the nearest neighbors, and 
$G_{ij}=\langle \boldsymbol{\Psi}_{MF}| \mathbf{S}_{i} \cdot \mathbf{S}_{j} | \boldsymbol{\Psi}_{MF}\rangle$. 
The notation in this expression implicitly assumes that these links are counted twice 
i.e. once for $k,l$ and the other for $l,k$, hence the factor of $2$ in the denominator. 

To obtain the optimal $w_i$, we define a quadratic 
cost function $C_{SMA}$,
\begin{equation}
C_{SMA} \equiv \frac{-\sum_{\langle k,l \rangle}(w_{k}-w_{l})^{2}G_{kl} }{2}- \Lambda \left(\sum_{i,j=1}^{N_{s}}w_{i}w_{j}G_{ij}-1   \right)
\label{eq:smacost}
\end{equation}
where $\Lambda$ is a Lagrange multiplier and is exactly equal to the SMA gap $\Delta_{SMA}$.
On differentiating $C_{SMA}$ with respect to $\{ w_{i}\}$ and setting the derivatives to zero, one gets a set of 
linear equations, which is compactly written as,
\begin{equation}
	\boldsymbol{\text{M}} \boldsymbol{\text{w}} = 2 \Lambda \boldsymbol{\text{G}} \boldsymbol{\text{w}}
\label{eq:gen_eig}
\end{equation}
where $\boldsymbol{\text{w}}$ is a compact 
notation for the vector $\{ w_i \}$ and  ${\text{G}}$ 
denotes the matrix of spin-spin correlations with 
entries $G_{ij}$. $\boldsymbol{\text{M}}$ is a matrix with entries given by,
\begin{subequations}
\begin{eqnarray}
	M_{ii} & = & + 2 \sum_{\langle j \rangle} G_{ij} \\
	M_{ij} & = & - 2 G_{ij} \;\;\;\;\;\; \text{for i,j connected } \\
        M_{ij} & = & \;0        \;\;\;\;\;\;\;\;\; \text{otherwise}
\end{eqnarray}
\end{subequations}
where $\langle j \rangle$ refers to the set of sites $j$ 
connected to site $i$. The generalized eigenproblem~\eqref{eq:gen_eig} is solved 
and the minimum eigenvalue yields the SMA gap.

\begin{figure}[htpb]
\centering
\includegraphics[width=\linewidth]{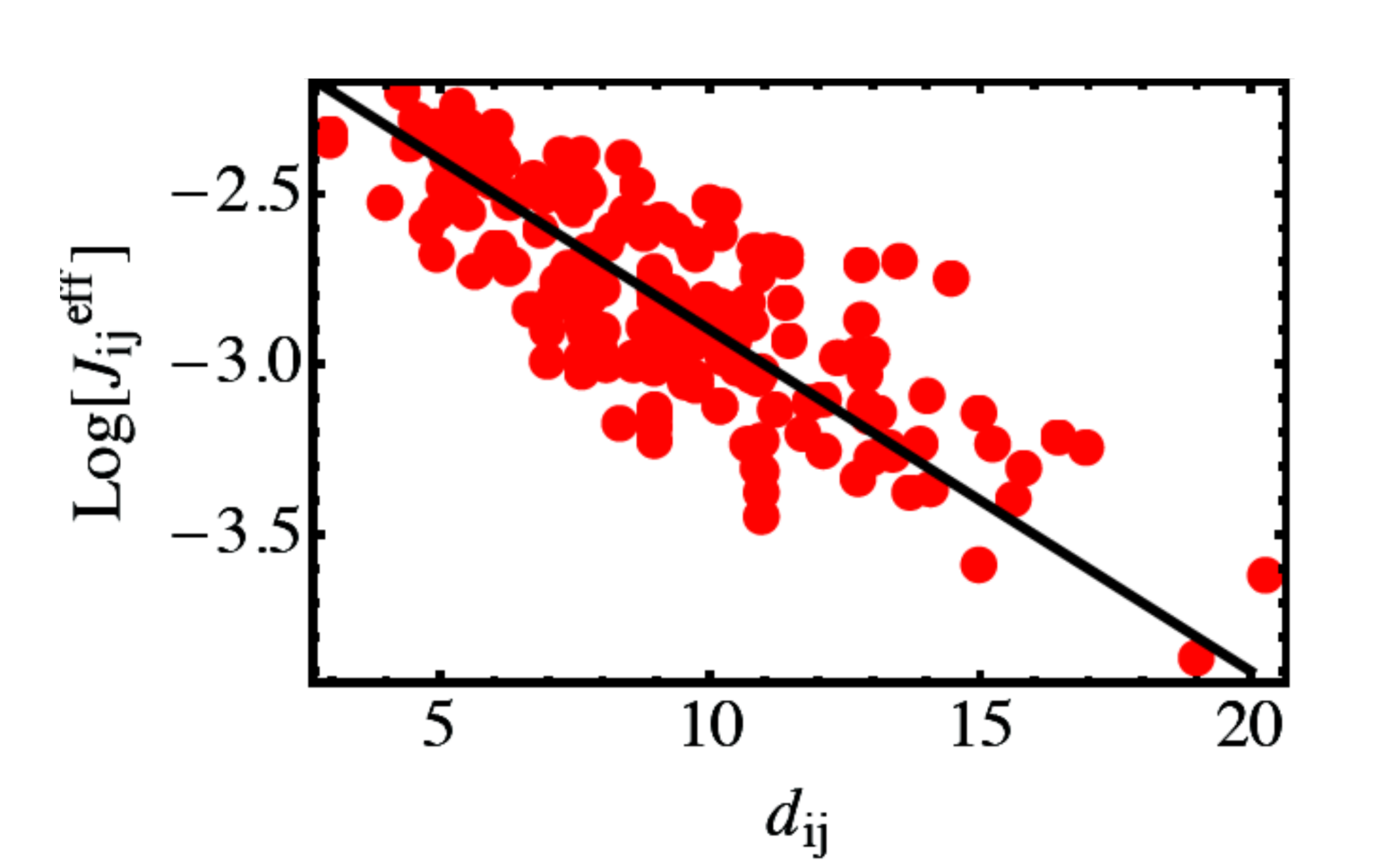} 
\caption{(Color online) 
The decay of effective interactions $J_{ij}^{eff}$ versus an effective distance 
$\tilde d_{ij}$ (see text). All data is for $N_{s}=50$ sized clusters. 
}
\label{fig:fig5}  
\end{figure}

The optimized SMA gap provides an upper bound for the effective 
interactions between two dangling spins on opposite sublattices. 
Since we desire the distance-dependence of effective 
couplings for extended objects, we define an "effective distance", 
analogous to that in Ref.~\onlinecite{changlani-ghosh}, 
$\tilde d_{ij}=\sum_{n=1}^{n_{d}}\sum_{i,j}\psi_{in}\psi_{jn}d_{ij}$ where $\psi_{in}$ 
is the amplitude of the single particle SBMFT mode $n$ at site $i$ and $d_{ij}$ is the distance between sites $(i,j)$. 
The sum over $(i,j)$ runs over all pairs of sites that the two spins can delocalize over.

We select percolation clusters from a randomly generated ensemble that 
have only two dangling spins on opposite sublattices. As is shown in Fig.~\ref{fig:fig5}, 
we find their effective interactions to decay exponentially; 
a fit to $J_{ij}^{eff}=J_{0}^{*}e^{-\tilde d_{ij}/\xi^{*}}$ gives $(J_{0}^{*},\xi^{*})=(0.15(2),10.1(1)))$. 
The decay length $\xi^{*}$ is an upper-bound on the decay constant $\xi \sim 5$ 
obtained from our previous DMRG study~\cite{changlani-ghosh}. This slow 
decay of the interactions with distance 
is due to the inability of the SMA to describe very small gaps, 
a situation that occurs when the number of dangling spins is large. 
\newt{However, the qualitative prediction} of exponentially decaying 
interactions is consistent with the occurence of localized SBMFT 
modes associated with each dangling spin~(region) 
as shown in Fig.~\ref{fig:fig1}.

\section{Conclusion} 
In summary, we have carried out 
Schwinger Boson mean field theory (SBMFT) 
calculations for the case of nonuniform geometries, with 
specific emphasis on percolation clusters on the square and Bethe lattice. 
We show how the theory predicts the formation of 
emergent spin degrees of freedom arising due to local 
sublattice imbalance~\cite{wangandsandvik, changlani-ghosh}.

Our approach involved an interpretation of the mean-field parameters, 
$\lambda_i$ and $Q_{ij}$, which were the 
on-site potential and bond-pairing amplitude respectively. 
We also showed that the low-lying single particle wave functions 
have their largest amplitudes in regions associated with sublattice imbalance 
(i.e. the "dangling spins"). Thus, these modes provide a way of detecting 
emergent degrees of freedom on percolation clusters.

This interpretation is made firm based on the observation 
that the number of low lying single particle frequencies correspond to the number 
of dangling spins on the cluster. The violations occur because 
the localized modes are not completely decoupled;
interactions between them further split the single particle energies. 
We generically found an additional lowering of the two lowest 
frequencies from this set; these were identified as the equivalent of (nonuniform) 
Goldstone modes. The fact that regions of sublattice imbalance are involved in these modes 
provides evidence for the link between the occurence of emergent degrees of freedom 
and long range order on the cluster, previously established numerically~\cite{sandvik2002}.

We explored how anharmonic effects in spin-wave theory 
may explain the lowest modes seen in SBMFT; after all, 
the LSWT Hamiltonian maps \emph{exactly} to a SBMFT Hamiltonian for 
large spin. In LSWT, the parameters $\lambda_i$ are fixed
by the coordination of the respective site $i$; on the other hand, 
in SBMFT they are variational parameters, which allows the lowest energy modes 
to have nonuniform amplitudes with large weights in dangling regions. 
Evidence for the \emph{tendency} to form localized modes for spin-1/2 is 
seen by going beyond LSWT i.e. to order $1/S$ using self-consistent Hartree Fock methods. 

These observations also motivated a preliminary exploration of the 
role of spin-length for the Heisenberg model on percolation clusters. 
Unlike the spin-1/2 case, we found our spin wave Hartree Fock results 
to converge for the spin-1 case, which we take as evidence of the reduced role of spin fluctuations. 
Exact diagonalization calculations on small clusters also suggest that 
the picture of "emergent localized spins" may no longer apply for spin-1 
as the distinction between the quasidegenerate states and the rest of the spectrum 
is not as clear as the spin-1/2 case. This hints at the increased role of the bulk spins 
in the low energy spectrum, expected of a spatially extended collective excitation.

Finally we comment that SBMFT for disordered systems 
provides reasonable qualitative insights, complementing other highly accurate many-body calculations 
such as DMRG. We expect our implementation of SBMFT for nonuniform situations to 
perform equally well and scale favorably even in other dimensions. 
Based on results presented here and ongoing work, we believe the theory will be a 
useful tool in the treatment of frustrated lattices with disorder. In addition, in 
interesting cases like that of the $Z_{2}$ spin liquid on the kagome lattice, 
one can implement modifications to the theory that create excitations (visons), 
leading to numerical realizations of topological excitations~\cite{lawler}.

\section{Acknowledgement} We thank Anders Sandvik, Daniel Arovas, Assa Auerbach, Michael Lawler, 
Ribhu Kaul, Sumiran Pujari and Victor Chua for discussions. 
S.G. and C.L.H. acknowledge support from the National Science Foundation 
under grant number NSF-DMR 1005466. S.G. was also supported by a Cornell Graduate 
Fellowship. H.J.C. was supported by the SciDAC program of the  
U.S. Department  of Energy (DOE) under Award Number DE-FG02-12ER46875. 
We acknowledge the Taub campus cluster at the University of Illinois at 
Urbana-Champaign and the CCMR facilities at Cornell University for computing resources.

\end{document}